\documentclass[10pt, a4paper,twocolumn]{article}

\usepackage{graphicx}
\usepackage{commath}
\usepackage{amsthm}
\usepackage{hyperref}
\usepackage{color}
\usepackage[table]{xcolor}
\usepackage{multirow}
\usepackage{booktabs}
\usepackage{subfigure}
\usepackage{float}
\usepackage{authblk}
\usepackage{qtree}
\usepackage[all,cmtip]{xy}

\usepackage[english]{babel}
\usepackage{amssymb,amsmath,amsthm, amsfonts}

\newtheorem{definition}{Definition}[section]

\RequirePackage[left=2cm,%
                right=2cm,%
                top=2.25cm,%
                bottom=2.25cm,%
                headheight=12pt,%
                letterpaper]{geometry}%

\title{Patterns on logistic map: back to front}

\author[1]{Jo\~ao Valle}
\author[1,*]{Odemir M. Bruno}
\affil[1]{\small{S\~{a}o Carlos Institute of Physics, University of S\~{a}o Paulo (USP), PO Box 369, 13560-970, S\~{a}o Carlos, SP, Brazil. \protect\\Scientific Computing Group - http://scg.ifsc.usp.br}}
\affil[*]{\small{bruno@ifsc.usp.br}}

\date{}
\begin{document}
\maketitle

\begin{abstract}


    
    Recently patterns were found in the least significant digits in the logistic map orbits in the chaotic regime. However, the dynamic of these digits was not explored in deep.
    We propose a new interpretation of the patterns found in the least significant digits of the logistic map in a chaotic regime, based on exploring the digits' dynamics.
    We explored these digits by the dynamic iteration method for low positions.
    When the dynamic of the least significant digits was explored, periodic and fixed digits were found with dynamic analogs to fixed and periodic points of discrete systems. Periodic digits designated well-established period sizes, which are independent of parameter control or initial condition. Fixed digits have attraction and stability characteristics, such as fixed points.
    This analysis demonstrates that the logistic map's least significant digits have characteristic dynamics.
\end{abstract}




\section{Introduction}
\label{sec:intro}

Although we have known about chaotic systems since the 19th century, especially through the works of Poincaré~\cite{barrow1997poincare}, it was only with the advent of computers that chaos theory could be consolidated~\cite{lorenz1963deterministic, mandelbrot1982fractal}. A large part of its conception is due to the study of its patterns, especially the visualization methods, such as bifurcation diagram~\cite{jafari2021simple}, Poincaré diagram, histograms, etc.; and modern machine learning methods~\cite{Ott_Machine_Learning, pathak2017using} which allowed a better understanding of its complex behaviors. Although much has been studied, many of the complex patterns and structures that arise in simple chaotic systems still intrigue us.

With the use of computers to understand chaos, the problem of limited accuracy arose. This problem causes the difference between the theoretical orbits to be different from those calculated in the computer after some iterations~\cite{nagaraj2008increasing, chia1991maps, grebogi1990shadowing}. Furthermore, precision influences the behavior of orbits, as seen in Lorenz's famous experiment for climate forecasting~\cite{lorenz1963deterministic} and, more recently, in the works~\cite{lai1997characterization, binder1986simulating, grebogi1988roundoff, grebogi1990shadowing, adler2001chaos}. Due to the need for high accuracies to calculate a chaotic orbit, the least significant digits are usually excluded in the truncation process. The analysis of these digits is rare, one of the few works that explore the least significant digits is in the Liu, J. et al. paper~\cite{liu2014property}.

Our contribution was to reveal the dynamics of the pattern digits shown in the Liu work~\cite{liu2014property}. We propose a new interpretation for these patterns, in which shows an analogy with the fixed and periodic points of the logistic map of global dynamics. In the broadest sense, we studied the complexity of the last digits of the logistic map orbits in a chaotic regime, the reverse path of the work~\cite{machicao2017improving, machicao2020zooming} made by our research group.

To explore the least significant numbers, we applied the iterative method for low positions proposed by Liu, J. et. al~\cite{liu2014property}. This method allows us to explore the dynamic without a long precision requirement. To investigate the dynamic properties, the cycle detection algorithm and the frequency distribution were needed.

Through pattern analysis, we observed the presence of periodic and fixed digits. These digits have a dynamic analogous to fixed and periodic points. Moreover, the dynamics of fixed digits present the attraction and stability phenomenon similar to the dynamics of fixed points. Periodic points have well-defined periods for each decimal position and cycle numbers appear with well-defined average frequencies.



To understand the patterns and analysis of the last digits of the logistic map, we have segmented the article into the following parts: a brief introduction to dynamical systems in Sec.~\ref{sec:dynamical_system}, description of the logistic map in Sec.~\ref{sec:logistic}, the fixed digit pattern in Sec.~\ref{sec:fixed}, the periodic digit pattern in Sec~\ref{sec:periodic}, the discussion in Sec.~\ref{sec:discussions} and conclusion in Sec.~\ref{sec:conclusion}.

\section{Dynamical System}
\label{sec:dynamical_system}

    A dynamical system is a system that evolves in time from a deterministic rule~\cite{strogatz2018nonlinear}. This evolution occurs in continuous or discrete time, which is called continuous or discrete dynamical systems, respectively. In this article, the discrete dynamical system is approached, in which its mathematical representation is given by Eq.~\eqref{eq:discret_system}.
    
    \begin{equation}
        x(n+1) = f(x(n))
        \label{eq:discret_system}
    \end{equation}
    
    or 
    
    \begin{equation}
        x_{n+1} = f(x_n)
        \label{eq:discret_system2}
    \end{equation}

    To understand the dynamics of Eq.~\eqref{eq:discret_system}, it is needed to perform the iteration process. This process is given by composing the equation repeatedly from a seed (initial condition). Through this process, it is possible to form the orbit (Eq.~\eqref{eq:orbit}) and classify the dynamics of the system. 
    
    \begin{equation}
        \mathcal{O}(x_0) = \{x_0, f(x_0), f^2(x_0), f^3(x_0), \ldots \}
        \label{eq:orbit}
    \end{equation}
    
    \noindent where $x_0 = x(0)$ is the seed and $f^2 = f \circ f$, $f^3 = f \circ f \circ f$, etc.

    After the transient period, formed by the first iterations of the dynamic system, the dynamics of $f$ is classified as constant, periodic and aperiodic and non-constant dynamics. The constant dynamics is formed by points in orbit that have the same value over the iterations. For periodic dynamics, the orbit is formed by points that presents a set of different values that are repeated after some iterations. For aperiodic and non-continuous dynamics, the points in orbit show a chaotic behavior.

    Even in chaotic or periodic behavior, the dynamics can become constant for certain seed values. Values capable of performing such behavior are called fixed points or equilibrium points. In some cases, the fixed point appears after a few iterations, being called an eventual fixed point. The fixed point is formally described by Def.~\ref{def:fixed}.
    
    
    \begin{definition}
    A point $x^*$ in domain of $f$ is fixed point of $ x_{n+1} = f(x_n)$ if $f(x^*) = x^*$.
    \label{def:fixed}
    \end{definition}
    
    Fixed points are classified according to their stability, the main classifications being: stable, unstable and attraction. According to~\cite{elaydi2007discrete}, the fixed point $x^*$ is said to be stable if, after $n$ iterations of the dynamic system given an initial condition $x_0$, the distance from the fixed point $x^*$ is sufficiently small . That is, $|f^n(x_0) - x^*| < \varepsilon$ for all $n \in \mathbb{Z}^+$, where $\varepsilon > 0$ and $|x_0 - x^*| < \delta$, with $\delta>0$. If $x^*$ does not follow this property, it is called unstable. The point $x^*$ is a fixed point of attraction if, after an extremely large number of iterations, the displayed value is equal to $x^*$. That is, $\lim_{n \longrightarrow \infty} f^n(x_0) = x^*$, for $|x_0 - x^*| < \eta$ with $\eta > 0$.
    
    
        
        
    
    In addition to fixed points, there are periodic points, characterized by values that repeat after a certain number of iterations. That is, when the discrete dynamical system is initiated or eventually presents a periodic point, its dynamics presents closed cycles.
    
    According to~\cite{elaydi2007discrete}, the point $\Bar{x}$ is said to be a periodic point of the dynamical system $f$ with period $k$, if $f^k(\Bar{x}) = \Bar{x }$, with $k \in \mathbb{Z}^+$. We can say that $\Bar{x}$ is a fixed point on the map $f^k$. If the point $\Bar{x}$ needs $m$ iterations to present the fixed point, we say that it is an eventually periodic point. That is, $f^{k+m}(\Bar{x}) = f^m(\Bar{x})$ for $k, m \in \mathbb{Z}^+$.
    
    
        
    
    
    Periodic points are present in chaotic regimes. Not only present, but they also need to be dense in the domain of the dynamic system $f$, according to Devaney's definition~\cite{devaneyChaos}. Among the countless definitions of chaos, the Def.~\ref{def:devaney}, described by Devaney, R., is the most well known.
    
    
    \begin{definition}
    A map $f:I \longrightarrow I$, where $I$ is an interval, is said to be a chaotic if:
    \begin{enumerate}
        \item $f$ has sensitive dependence on initial conditions;
        \item The set of periodic points $P$ is dense in $I$;
        \item $f$ is transitive.
    \end{enumerate}
    \label{def:devaney}
    \end{definition}
    
    Sensitivity to initial conditions indicates that small perturbations in initial conditions lead to eventual divergences in their temporal evolution. Regarding the second characteristic, it indicates that any trajectory from a periodic point can approach the other periodic points in the phase space. Regarding transitivity, it implies that the trajectory initialized at any point in the phase space can eventually cover the entire phase space when evolving temporally.


\section{Logistic map}
\label{sec:logistic}
	
    

    The logistic map is a discrete dynamical system known to exhibit chaotic behavior, described by Eq.~\eqref{eq:logistic_map}.
    This equation emerged through the study of population dynamics, through the discretization of the logistic equation using Euler's method.
    Being popularized by the article by May~\cite{logisticMap}, where the variety of dynamics presented by a simple mathematical model, the logistic map, was shown.
    
    \begin{equation}
        f(x_n) = x_{n+1} = \mu x_n (1 - x_n)
        \label{eq:logistic_map}
    \end{equation}
    
    \noindent where $f : [0,1] \longrightarrow [0,1]$ and $\mu \in (0, 4]$.
    
    The logistic map displays several dynamics according to the value of the control parameter $\mu$. Covering periodic, constant and chaotic dynamics. In addition, $\mu$ also changes the quantity and stability of fixed points and the existence of periodic points.
    For example, $x^*_1 = 0$ and $x^*_2 = 3/4$ are fixed points of the logistic map with $\mu = 4$, obeying Def.~\ref{def:fixed}. For $\mu = 3.2$, the logistic map presents periodic points of period $2$ given by $\Bar{x}_1 \approx 0.513 $ and $\Bar{x}_2 = 0.799$.

    

    
    The logistic map presents fixed and periodic values not only in the values of the orbits, but also in the digits of these orbits. The work by Liu, J. et al in \cite{liu2014property} shows this behavior in the logistic map digits. It is possible to observe the presence of periodic and fixed digits even in a chaotic regime.
    
    
    For better understanding, Eq.~\eqref{eq:decimal_expansion} represents the value $x$ of the orbit in decimal expansion, where $d_n \in \{0,1,\ldots,9 \}$. Through this representation, Liu, J. et al explored the least significant digits, from $d_m$ to a certain $d_{m-p}$ with $p \in \mathbb{Z}^+$.

    \begin{equation}
        x = \sum_{n=1}^m\frac{d_n}{10^n} = 0.d_1d_2d_3\ldots d_m
        \label{eq:decimal_expansion}
    \end{equation}

    To numerically explore these digits, Liu, J. et al developed the dynamic iteration method for low positions. This method consists of always using a fixed number of decimal places, excluding $k$ digits after the decimal separator to keep this fixed number of decimal places. Using this approach, the use of libraries of multiple precision becomes useless and the results become more reliable.
    
    
    
    
    
    
    In this article we will continue the work developed by Liu, J. et al, giving a new interpretation and exploring the properties of the least significant digits of the logistic map. In the next two sections, the results of this new approach are presented.

\section{Fixed digits}
\label{sec:fixed}

    In this section, the fixed digits that occur in logistic map iterations was explored. These fixed digits appear in the least significant digits of the logistic map's orbit calculated without truncation. For a better understanding, the fixed digits of a specific case ($\mu = 4.0$) will explained in the first part and, in the second part, the more general result is shown.
    
    \subsection{Case: $\mu = 4.0$}
        
        When calculating the first iterations of the logistic map, the fixed digits appear in the last decimal places. For instance, choosing the initial value $x_0 = 0.01$, the number $6$ is always present in the last decimal place throughout the iterations, as seen below:
        
        \begin{equation*}
            \begin{split}
                f(x_0) &= 0.039\textbf{6}\\
                f^2(x_0) &= 0.1521273\textbf{6}\\
                f^3(x_0) &= 0.515938505357721\textbf{6}\\
                f^4(x_0) &= 0.9989838561878475194064770275737\textbf{6}\\
                \ldots
            \end{split}
        \end{equation*}
        
        The goal in this section is to find other fixed numbers besides $6$ for the case of $\mu = 4.0$ and explore their properties. Also show how the digits behave similar to the fixed points of discrete dynamical systems.
        
        \subsubsection{Finding the fixed numbers}
        
        The procedure for finding fixed digit numbers is analogous to fixed points, described by Def.~\ref{def:fixed}. Fixed digits are found by comparing the last digits that have the same value after an iteration. This procedure was performed numerically.
        
        The numerical method for finding the fixed digits consists of comparing the last $k$ digits of an initial condition $x_0$ and its first iteration $f(x_0)$. Follow the steps of the method:
        
        \begin{enumerate}
            \item For fixed digit of one decimal place, test the seeds $x_0 \in \{0.1, 0.2, 0.3, 0.4, 0.5, 0.6, 0.7, 0.8, 0.9 \}$. The last digit of the seed is compared with the last digit of the first iteration $f(x_0)$. The seeds that have the same value have a fixed number, named $x_{f_1}$.
            
            \item For the fixed number to two decimal places, use $x_{f_1}$ from the previous step and $d \in \{0,1,2,3,4,5,6,7,8,9 \}$. The new seed is defined by $x_0 = (x_{f_1} + d)/10$. When comparing the last two digits of $x_0$ with the last two digits of $f(x_0)$, the $x_{f_2}$ is defined by $x_0$ which had equal digit values in the comparison. 
            
            \item For the fixed number of $m$ decimal places, use the $x_{f_{m-1}}$ from step $m-1$ and the same $d \in \{0,1,2,3,4,5,6,7 ,8,9 \}$. The seed is defined in the same way: $x_0 = (x_{f_{m-1}} + d)/10$. And the new $x_{f_m}$ is given by the $x_0$ which has the last $m$ digits equal to the last $m$ digits of $f(x_0)$.
        \end{enumerate}
        

        For the case $\mu = 4.0$, the fixed numbers with one decimal place are: $6$ and $5$. When performing the procedure described above, the class of fixed numbers arising from $6$ grows indefinitely and the class arising from $5$ forks into two other classes of fixed numbers: $975$ and $375$, both of which also grow indefinitely. Therefore, for $\mu = 4.0$, we have three varieties of fixed numbers, as show below:

        \Tree[.$6$ [.$56$ [.$656$ [.$2656$ [.$22656$ [.$222656$ [.$\cdots$ ] ] ] ] ] ] ] \Tree[.$5$ [.$75$ [.$375$ [.$4375$ [.$34375$ [.$734375$ [.$\cdots$ ] ] ] ] ][.$975$ [.$9975$ [.$99975$ [.$999975$ [.$\cdots$ ] ] ] ] ]]]

        The three classes of fixed digits have different properties. This property is called fixed digit stability, as described in the next subsection.
        
        \subsubsection{Stability}
        
        These three classes of fixed numbers ($6$, $975$ and $375$) exhibit two stability: stable or expansive. The stable behavior is indicated by the permanence of a fixed number with a certain number of decimal places fixed throughout the iterations of the logistic map. Meanwhile, the expansive behavior is  the increase in the number of decimal places occupied by the fixed digits over the iteration of the logistic map, this increase occurs monotonically.
        
        Classes $6$ and $975$ have stable behavior, that is, the number of decimal places of the fixed number remains constant throughout the iterations, after a transition period. In some cases, the number of places expands to a next fixed number with more decimal places, and in the next iterations this number of decimal places remains stable constant. For example, the case of the fixed digits $656$ that passes to $2656$:

        \begin{equation*}
            \begin{split}
                f(x_0) &= 0.\textbf{656}\\
                f^2(x_0)& = 0.90\textbf{2656}\\
                f^3(x_0)& = 0.\cdots 7258\textbf{2656}\\
                f^4(x_0)& = 0.\cdots 9194\textbf{2656}\\
                f^5(x_0)& = 0.\cdots 3066\textbf{2656}\\
                f^6(x_0)& = 0.\cdots 0810\textbf{2656}\\
                \cdots
            \end{split}
        \end{equation*}
        
        The $375$ class has an expansive behavior, that is, the number of decimal places of the fixed number increases monotonically in each iteration. This behavior is explained by the attraction of fixed numbers, in which the fixed number with the lowest number of decimal places falls in the region of attraction of the fixed number with the highest number of decimal places, as show in the next subsection. The expansive behavior is shown in the example below:
        
        \begin{equation*}
            \begin{split}
                f(x_0)& = 0.9\textbf{375}\\
                f^2(x_0)& = 0.2\textbf{34375}\\
                f^3(x_0)& = 0.71\textbf{77734375}\\
                f^4(x_0)& = 0.\cdots 98919\textbf{677734375}\\
                f^5(x_0)& = 0.\cdots 02709579\textbf{4677734375}\\
                f^6(x_0)& = 0.\cdots 824606895\textbf{44677734375}\\
                \cdots
            \end{split}
        \end{equation*}
        
        The classes resulting from the bifurcation ($975$ and $375$) present two distinct stabilities, even having the last two digits in common.
        
        \subsubsection{Attraction}
        
        The attraction phenomenon does not only exist for the fixed points in the global dynamics of the logistic map, but also exists in the logistic map digits. It is possible to determine which seed is attracted by a certain fixed digit after some iterations. For example, to determine which seed presents certain fixed number after one iteration, just substitute correctly the numbers in Eq.~\eqref{eq:attraction} and find its integer solutions.
        
        \begin{equation}
            4 x (10^k - x) \mod 10^k = x_f
            \label{eq:attraction}
        \end{equation}
        
        \noindent where $x, k, x_f \in \mathbb{Z}^+$.
        
        In the Eq.~\eqref{eq:attraction}, the terms are positive integers. For example, for fixed digits of two decimal places, such as $56$, the terms of the equation are: $x_f = 56$ and $k=2$, where $x_f$ represents the fixed digit and $k$ the number of places decimals. While the $x$ represents the value of the last $k$ decimal places of the seeds that have the fixed digit $x_f$ in the last $k$ decimal places after one iteration. The integer solutions of the Eq.~\eqref{eq:attraction} are the possible values of $x$. Using $x_f=56$ and $k=2$, the Eq.~\eqref{eq:attraction} becomes as shown in Eq.~\eqref{eq:example56}.
        
        
        \begin{equation}
            4 x (100 - x) \mod 100 = 56 
            \label{eq:example56}
        \end{equation}
        
        
        
        
        The integer solutions of the Eq.~\eqref{eq:example56} between $0$ and $100$ are the possible values of $x$: $94, 06, 56, 44, 19, 81, 69, 31$. These numbers represent the last two digits a seed needs to have to pull to digit $56$ after one iteration. Then, there are $8$ different ending numbers that are attracted by $56$ after one iteration.
        
        
        However, some seeds need two iterations to be attracted by the fixed digit $56$. The first iteration of these seeds presents the last two digits equal to the complement of $56$, the number $44$, and, only in the second iteration, presents the end $56$. To determine the last two possible digits of the seeds that show this behavior, find the integer solutions of $x$ with $k$ digits in the second iteration of Eq.\eqref{eq:attraction}. Using the same values, $x_f = 56$ and $k=2$, the Eq.~\eqref{eq:attraction2} represents this kind of attraction.
        
        \begin{equation}
            4 (4 x (100 - x)) (100 - (4 x (100 - x))) \mod 100 = 56 
            \label{eq:attraction2}
        \end{equation}
        
        
        
        
        The integer solutions with $2$ digits of $x$ are: $08, 44, 56, 92, 06, 42, 58, 94, 17, 19, 31, 33, 67, 69, 81, 83$. Removing the numbers that already appeared in the previous solution, the restricted solutions are given by: $08, 92, 42, 58, 17, 83, 33, 67$. Hence, seeds that end with the last two digits equal to some value of the restricted solution need two iterations to be attracted by the fixed digit $56$.
        The behavior of the two last digits through iterations of these seeds is given by:
        
        \begin{equation*}
            44 \longrightarrow 56 \longrightarrow 56 \longrightarrow 56 \longrightarrow \ldots
        \end{equation*}
        
        In the numerical experiments performed, no seeds were found that need more than two iterations to be attracted by some fixed digit. Furthermore, expansive stability of the fixed digit $375$ can be justified by attracting a fixed digit with a lower decimal place by one with a higher decimal place.

    \subsection{General Case}
    
        For the other values of $\mu$, there is also the presence of three classes of fixed digits and, for very specific cases, there is the presence of a class of fixed digits.
        
        \subsubsection{$1$-class case}
            
            The values of $\mu$ that have a class of fixed digits follow a pattern. This pattern occurs with some $\mu$ ending in $e$, where $e = \{ 2, 4, 6, 8 \}$. Considering the last two digits of $\mu$, defined as $d_2d_1$, where $d_1 \in e$, if $\mu$ is the $1$-class case, the next $\mu$ with $1$-class case, the $d_2$ incremented by one decimal unit and $d_1$ follows the cycle shown below:
            
            \begin{align*}
              \xymatrix{ 
               2 \ar@(r,u)[r]	
               & 8 \ar@(r,u)[r] 
               & 6 \ar@(r,u)[r] 
               & 4 \ar@(l,d)[lll]\\
              }
            \end{align*}
            
            For example, the set of control parameters between $3.911$ and $3.999$, with the same number of decimal places and belongs to the $1$-class case, have the last digit $d_1$ belonging to the cycle described above while the penultimate digit $d_2$ is incremented a decimal unit. The values of $\mu$ classified as $1$-class in the range $[3.911, 3.999]$ are listed below:
    
            \begin{equation*}
                    \begin{split}
                        3.91\textbf{2}\\
                        3.92\textbf{8}\\
                        3.93\textbf{6}\\
                        3.94\textbf{4}\\
                        3.95\textbf{2}\\
                        3.96\textbf{8}\\
                        3.97\textbf{6}\\
                        3.98\textbf{4}\\
                        3.99\textbf{2}
                    \end{split}
            \end{equation*}
            
            This cycle occurs for the other $\mu$ values with the cycle of the last digit $d_1$ started by another value in the set: $\{ 2, 4, 6, 8 \}$. For example, in the case of parameters $3.9811$ to $3.9899$, the cycle starts at $3.9816$ then moves to $3.9824$, following the cycle.
            
        \subsubsection{$3$-classes case}
        
            The control parameter $\mu$ ending with numbers other than $1, 3, 5, 7$ or $9$ always has $3$ fixed digits classes. Control parameters ending in $2, 4, 6$ or $8$ have only one fixed number class if they follow a rule shown in $1$-class case, otherwise the control parameter has $3$ fixed digits classes.
            
            Some of these classes arise by the bifurcation phenomenon, as seen with $\mu = 4.0$, with the $375$ and $975$ classes. The parameters ending with numbers: $2, 4, 5, 6$ or $8$ show the bifurcation phenomenon, although control parameters ending with $1, 3, 7$ or $9$ do not show the bifurcation phenomenon. For example, for $\mu = 3.9916$, that ending with $6$, the two classes come from a bifurcation, while one class not, as shown below:

            \Tree[.$4$ [.$64$ [.$064$ [.$6064$ [.$96064$ [.$496064$ [.$\cdots$ ] ] ] ] ] ] ] \Tree[.$5$ [.$25$ [.$125$ [.$3125$ [.$53125$ [.$453125$ [.$\cdots$ ] ] ] ] ][.$525$ [.$9525$ [.$59525$ [.$059525$ [.$\cdots$ ] ] ] ] ]]]
                
                
            
            For $\mu = 3.9913$, that ending with $3$, there is no bifurcation phenomenon, as shown below:
            
            \Tree[.$3$ [.$23$ [.$023$ [.$4023$ [.$14023$ [.$114023$ [.$\cdots$ ] ] ] ] ] ] ] \Tree[.$5$ [.$75$ [.$375$ [.$4375$ [.$34375$ [.$734375$ [.$\cdots$ ] ] ] ] ] ] ] \Tree[.$8$ [.$48$ [.$648$ [.$9648$ [.$79648$ [.$379648$ [.$\cdots$ ] ] ] ] ] ] ]

            Furthermore, the control parameters $\mu$ with three classes are more frequent than the parameter with one class. Therefore, the $3$-class case is more likely to be found than the $1$-class case.
    
    \subsubsection{Stability}
    
        For the $\mu$ classified as $1$-class case, the fixed digits stability is always stable. However, for $\mu$ classified as $3$-classes case, one class has expansive behavior and the other two classes are stable, regardless of whether the fixed digit class arises from the bifurcation. Furthermore, the expansive stability classes always occur with fixed numbers ending in $5$.
    
    \subsubsection{Attraction}
        
        As described for the case $\mu = 4$, the fixed digits show the phenomenon of attraction. A seed with certain values in the last digits displays after one or two iterations display a certain fixed digits in the last decimal places. With the integer solutions of the Eq.~\eqref{eq:attraction_mu}, it is possible to determine all possible last digits of the seed that will be attracted to the fixed digit. Where the Eq.~\eqref{eq:attraction_mu} is a generalization of the Eq.~\eqref{eq:attraction}.
        
        \begin{equation}
                \tilde{\mu} x (10^k - x) \mod 10^k = x_f
                \label{eq:attraction_mu}
        \end{equation}
        
        \noindent where $\tilde{\mu},x, k, x_f \in \mathbb{Z}^+$. 
        
        The terms of the equation are all positive integers. The $\tilde{\mu}$ is the control parameter in the integer form, that is, if $\mu = 3.829$, then $\tilde{\mu} = 3829$. The other parameters have the same meaning as Eq.~\eqref{eq:attraction}, where $x_f$ are the fixed digits of $k$ decimal places, $k$ is the decimal place number and $x$ are the last $k$ digits of the seed that will be attracted by $x_f$ after one iteration.
        
        If we substitute the values of $k,x_f$ and $\tilde{\mu}$ into the Eq.~\eqref{eq:attraction_mu}, the integer solutions of $x$ represent all possible $k$ last digits of the seed which will be attracted by the fixed digit $x_f$ after one iteration. Also, if we define $x=x_f$ and substitute the values of $\tilde{\mu}$ and $k$ into Eq.~\eqref{eq:attraction_mu}, the integer solutions of $x_f$ with $k$ digits represent all existing fixed digits with $k$ digits for the chosen $\mu$.

\section{Periodic numbers}
\label{sec:periodic}

    In this section, the periodic digits are described. The presence of these digits only occurs in orbits with stable fixed digits and the decimal places occupied by the periodic digits are the decimal places subsequent to the fixed digits. In the first part, the periodic digits for the case $\mu = 4$ is described, and then the general case of the periodic digits is shown.
    
    \subsection{Case: $\mu = 4.0$}
        
        As the periodic digits only present in orbits with stable fixed digits. Therefore, for the case $\mu = 4$, only the $6$ and $975$ classes have the periodic digits. After passing through the transition period and the fixed digits occupy the last $k$ decimal places constantly, the periodic digits occupy the remaining decimal places.
    
        For example, for $x_0 = 0.01$ and $\mu = 4.0$, the first $8$ iterations have the following least significant numbers:
        
        \begin{equation*}
            \begin{split}
                f(x_0)& = 0.\ldots \textbf{9}6\\
                f^2(x_0)& = 0. \ldots 36\\
                f^3(x_0)& = 0. \ldots 16\\
                f^4(x_0)& = 0. \ldots 76\\
                f^5(x_0)& = 0. \ldots \textbf{9}6\\
                f^6(x_0)& = 0. \ldots 36\\
                f^7(x_0)& = 0. \ldots 16\\
                f^8(x_0)& = 0. \ldots 76\\
            \end{split}
        \end{equation*}
        
        In this one, the fixed digit is represented by the number $6$, which belongs to one of the stable classes present for $\mu=4$. Setting $l=1$ as the decimal place of the fixed digit, notice in the decimal place $l=2$, the periodic digits are present. These periodic digits have a period equal to $4$, following the cycle described below:
        
        \begin{align*}
          \xymatrix{ 
           9 \ar@(r,u)[r]	
           & 3 \ar@(r,u)[r] 
           & 1 \ar@(r,u)[r] 
           & 7 \ar@(l,d)[lll]\\
          }
        \end{align*}
        
        For the decimal place $l=3$, there are other periodic digits of period equal to $20$. The cycle that the periodic digits in $l=3$ is shown below:
        
        \begin{align*}
          \xymatrix{ 
            7 \ar@(r,u)[r] & 2 \ar@(r,u)[r] & 3 \ar@(r,u)[r] & 4 \ar@(r,u)[r] & 9 \ar@(r,u)[r] & 6 \ar@(r,u)[r] & 1 \ar@(r,u)[r] & 0 \ar@(r,u)[ld]\\ 7 \ar@(r,u)[d] & 4 \ar@(l,u)[l] & 3 \ar@(l,u)[l] & 6 \ar@(l,u)[l] & 9 \ar@(l,u)[l] & 0 \ar@(l,u)[l] & 1 \ar@(l,u)[l] \\ 2 \ar@(r,u)[r] & 5 \ar@(r,u)[r] & 8 \ar@(r,u)[r] & 5 \ar@(r,u)[r] & 8 \ar@(r,d)[lllluu]
            }
        \end{align*}
        
\begin{figure*}[!htb]
            \centering
            \includegraphics[width=\textwidth]{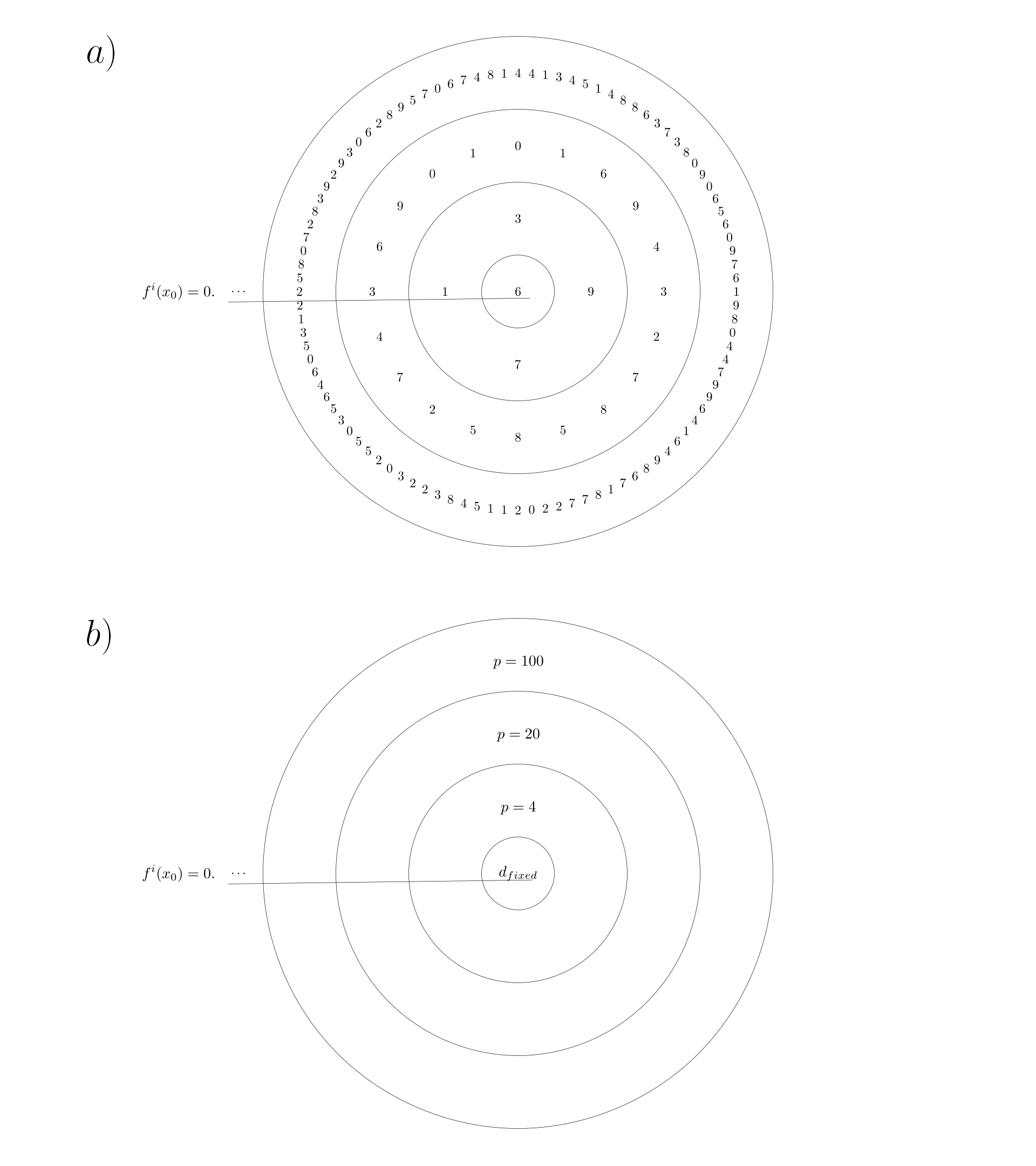}
            \caption{$a)$ schematic representation of the last digits of the logistic map, for $\mu=4$. The number $6$ is the stable fixed digit and the other disks are formed by periodic digits of period $4, 20$ and $100$, from the smallest disk to the largest. Each iteration represents a rotation on these disks. $b)$ schematic representation of the last digits of the logistic map. The $d_{fixed}$ represents the stable fixed digits, with $n$ decimal places, and the other disks are formed by periodic digits of period $4, 20$ and $100$, from the smallest disk to the largest. Each iteration represents a rotation on these disks.}
            \label{fig:periodScheme}
        \end{figure*}

        The periodic digits have increasing periods as $l$ increases. The period increases from a given decimal place $l$ is five times greater than the decimal place after it, for $l>2$. For example, for decimal place $l=4$, the cycle has period $100$; for $l=5$, the cycle has period $500$; so on. This characteristic is schematized by Fig.~\ref{fig:periodScheme}$a)$, where the central disk is represented by the fixed digit and the other disks are represented by the cycle of periodic numbers. The horizontal line represents the $i$-th iteration of the logistic map, given by $0.\cdots 2316$ in Fig.~\ref{fig:periodScheme}$a)$. The iterations are represented by the clockwise rotation of all disks simultaneously.
        
        For stable fixed digits with more decimal places, the behavior of the periodic digits occurs similarly to the previous example. The digit closest to the fixed digit has a period of $4$ and the other digits have a period multiple of $5$. The only difference is that these periodic digits occupy decimal places with higher $l$. For example, for $\mu =4$ and $x_0 = 0.56$, the least significant digits in the first $8$ iterations have the following behavior:
        
        \begin{equation*}
            \begin{split}
                f(x_0)& = 0.\ldots \textbf{8}56\\
                f^2(x_0)& = 0. \ldots 056\\
                f^3(x_0)& = 0. \ldots 456\\
                f^4(x_0)& = 0. \ldots 256\\
                f^5(x_0)& = 0. \ldots \textbf{8}56\\
                f^6(x_0)& = 0. \ldots 056\\
                f^7(x_0)& = 0. \ldots 456\\
                f^8(x_0)& = 0. \ldots 256\\
            \end{split}
        \end{equation*}
        
        In this case, the stable fixed digit occupies $2$ decimal places, present in the decimal place $l=1$, as the number $6$ and in the decimal place $l=2$, as the number $5$. The periodic digit only appears in the decimal place $l=3$, with a period equal to $4$, following a cycle as shown below:
        
        \begin{align*}
          \xymatrix{ 
           8 \ar@(r,u)[r]	
           & 0 \ar@(r,u)[r] 
           & 4 \ar@(r,u)[r] 
           & 2 \ar@(l,d)[lll]\\
          }
        \end{align*}
        
        For every seed that is attracted to a certain fixed digit, the periodic numbers always follow the same cycle. Changing the seed value only changes the cycle number that the periodic digit starts, a kind of lag. The cycle is changed only when the orbit has fixed digits that occupy more or less decimal places.

        \subsubsection{Frequency of cyclic digits}
        
\begin{figure}[!htb]
            \centering
            \includegraphics[scale = 0.5]{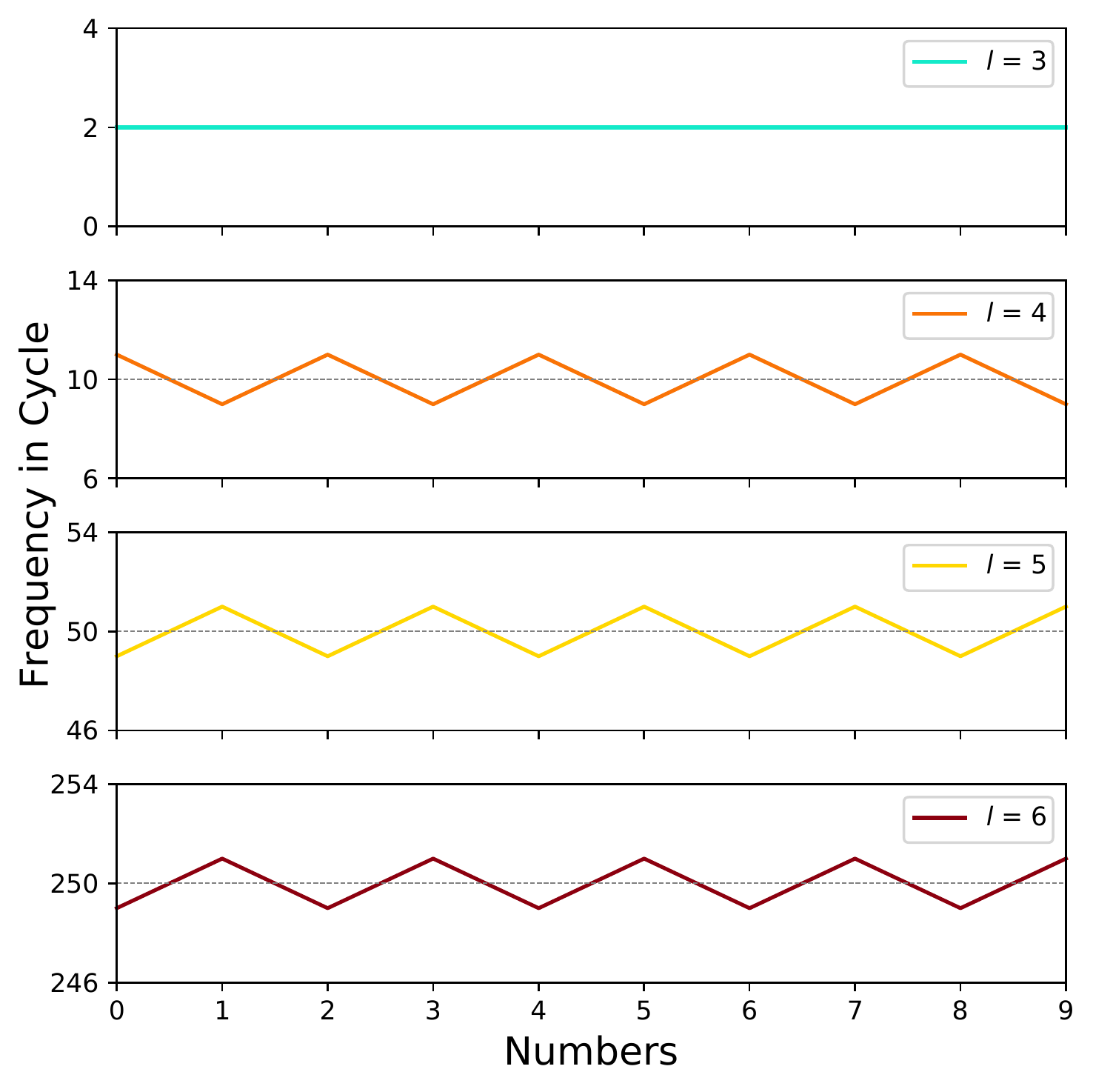}
            \caption{The plot of the frequency of each number belonging to the set $\{0,1,2,3,4,5,6,7,8,9\}$ that appears in position $l$, where $l \in {3,4,5,6}$, along the iteration of the logistic map with $\mu = 4$ and $x_0 = 0.1$.}
            \label{fig:frequencyL}
        \end{figure}

        For the periodic digits with period multiples of $5$, the frequency that the numbers $0, 1, 2, 3, 4, 5, 6, 7, 8$ and $9$ appear in the cycle is, on average, next equal to period divided size by $10$. For example, for $\mu=4$ and $x_0=0.01$, the numbers $0, 1, 2, 3, 4, 5, 6, 7, 8$ and $9$ appear twice in the $l = 3$, that is, each number appears with a frequency equal to the period $20$ divided by $10$. For other decimal places, the frequency that each number appears oscillates around the period value divided by $10$. The frequency to decimal places $l \in \{3,4,5,6\}$ is shown in the Fig.~\ref{fig:frequencyL}.
        
    \subsection{General case}

        For the general case, the same observations made for $\mu = 4.0$ are valid, such as a period cycle $4$ appears in the decimal place subsequent to the fixed digits and the other decimal places have period cycles multiple of $5$. Also, the frequency of numbers appearing in the $5$ multiple period cycle is, on average, equal to the period size divided by $10$.
        
        A schematic representation of how periodic and fixed digits form orbits is given by Fig.~\ref{fig:periodScheme}$b)$. In this figure, the central disk is formed by the stable fixed digits $d_{fixed}$ occupying $k$ decimal places. The other disks are formed by the cycles of the periodic digits in the decimal places subsequent to the decimal places of the fixed digits, the innermost disk being formed by the periodic cycle $p=4$, the second innermost formed by the period cycle $p=20 $, and so on. The horizontal line represents the orbit number of the $i$-th iteration and the rotation of all disks simultaneously represents the iteration of the logistic map $f$.
        
        
        The Eq.~\eqref{eq:period_size} shows a general way to calculate the period size $p$ for each $l$ position.
    
        \begin{equation}
            p = 
            \begin{cases}
            1 \text{ if } 1 \leq l \leq s_f\\
            4 \text{ if } l = s_f + 1 \\
            4 \cdot 5^{l-s_f-1} \text{ if } l > s_f + 1
            \end{cases}
            \label{eq:period_size}
        \end{equation}
        
        \noindent where $l$ is the decimal place, $s_f$ is the number of decimal places that the fixed digits occupy.
        
        The Eq.~\eqref{eq:frequency} shows a general way to calculate the average frequency of numbers appearing in the cycle for each cycle with periods multiples of $5$.
        
        \begin{equation}
            f = \frac{4 \cdot 5^{l-s_f-1}}{10}  \text{ for } l > s_f + 1
            \label{eq:frequency}
        \end{equation}
        
        Both the calculation of the period size by Eq.~\eqref{eq:period_size} and the calculation of the frequency of numbers $0, 1, 2, 3, 4, 5, 6, 7, 8$ and $9$ in the cycle of periodic digits by Eq.~\eqref{eq:frequency} are independent of the number of decimal places that the digit fixed occupies as long as it is a stable fixed digit.

\section{Discussions}
\label{sec:discussions}
    
    In this article, we saw that the logistic map shows patterns in the last digits, even in a chaotic regime. Furthermore, these patterns, formed by the fixed and periodic digits, present a dynamic similar to the global dynamics of the logistic map. Since the fixed digits are similar to fixed points, both have the same value over the iterations and present the phenomenon of attraction. Furthermore, fixed points and fixed digits can be classified according to their stability. In the case of the dynamics of dynamical systems, the stability of fixed points is classified as stable, unstable, or attraction; but in the case of digit dynamics, fixed digit stability is classified as stable or expansive.
    
    Another relationship between the dynamics of the logistic map and the dynamics of the digits of the logistic map is between the periodic points and the periodic digits. The periodic points have different periods depending on the value of $\mu$, however, the periodic digits have well-defined periods, and may have a period $4$ or a period multiple of $5$.
    
    The patterns presented in the last digits of the logistic map complemented with the \textit{deep-zoom} method, which shows that the absence of the first digits after the decimal separator makes the orbit more random-like. This is an indicator that the digits away from the decimal separator and from the last digits present high entropy. That is, the most central digits of the logistic map orbits in a chaotic regime tend to have greater entropy in relation to the digits at the extremes, given the patterns found in the regions.
    
    Periodic digits or expansive fixed digits are unlikely to reach the digits closest to the decimal point. This is because the number of decimal places always increases over iterations at a rate higher than the growth rate of the decimal places of the expanding fixed digit and the rate of emergence of new periodic digits, which need a transient to appear. 


    The results found in this article are valid only for rational numbers with finite decimal places. Only in this way can the last digits be visualized computationally. Mathematically speaking, rational numbers with finite decimal places are factored by the primes $2$ and $5$, that is, they are very sparse numbers within the domain of the logistic map. Therefore, the article deals with specific cases within the domain of the logistic map.

    
    



\section{Conclusion}
\label{sec:conclusion}


    Chaotic dynamical systems are known to be poorly predictable, with aperiodic and apparently random behavior. However, in this article, the logistic map, one of the most famous chaotic maps, presents patterns in the last digits in its orbit values, even in a chaotic regime. Furthermore, these patterns are given by fixed and periodic digits that have a dynamic similar to the fixed and periodic points of the global dynamical systems. Fixed digits also present the phenomenon of attraction also present in the global dynamics of dynamical systems. The periodic digits are similar to the periodic points, but the periodic digits have well-defined periods: period equals $4$ or a multiple of $5$. Therefore, in this article, we show that there is a clear dynamic of the last digits of the logistic map, a kind of \textit{subdynamics}.
    

\section*{Acknowledgments}
J. P. V. acknowledges support from S\~ao Paulo Research Foundation FAPESP (grant \#2022/01935-2).
O. M. B. acknowledges support from CNPq (grant \#307897/2018-4) and FAPESP (grant \#18/22214-6 and grant \#21/08325-2).


\end{document}